%% file: neurips_2021.tex
\documentclass{article}

\usepackage[final, nonatbib]{neurips_2021_ml4ps}

\usepackage[utf8]{inputenc} 
\usepackage[T1]{fontenc}    
\usepackage{hyperref}       
\usepackage{url}            
\usepackage{booktabs}       
\usepackage{amsfonts}       
\usepackage{nicefrac}       
\usepackage{microtype}      
\usepackage{xcolor}         

\input{math_commands.tex}

\usepackage{hyperref}
\usepackage{url}

\usepackage{subcaption}
\usepackage{caption}
\usepackage{cleveref}

\usepackage{graphicx}
\graphicspath{ {./images/} } 

\usepackage{amsfonts}
\usepackage{amsmath}

\usepackage{bm}  
\usepackage{mathtools}
\usepackage{dsfont}

\usepackage{courier}

\usepackage{multicol}

\usepackage{booktabs}
\usepackage{multirow}

\usepackage{enumitem}

\title{Can semi-supervised learning reduce the amount of manual labelling required for effective radio galaxy morphology classification?}

\author{Inigo V.~Slijepcevic \\
Department of Physics and Astronomy\\
University of Manchester, UK\\
inigo.slijepcevic@postgrad.manchester.ac.uk
\And
Anna M.~M.~Scaife\thanks{The Alan Turing Institute, 96 Euston Rd, London, UK \texttt{a.scaife@turing.ac.uk}} \\
Department of Physics and Astronomy\\
University of Manchester, UK\\
anna.scaife@manchester.ac.uk
}

\begin{document}

\maketitle

\begin{abstract}

In this work, we examine the robustness of state-of-the-art semi-supervised learning (SSL) algorithms when applied to morphological classification in modern radio astronomy. We test whether SSL can achieve performance comparable to the current supervised state of the art when using many fewer labelled data points and if these results generalise to using truly unlabelled data. We find that an artifical SSL scenario provides additional regularisation and improves baseline test set accuracy for a range of labelled data volumes. However, using truly unlabelled data degrades accuracy, which we show may be due to class imbalance in the unlabelled data.


\end{abstract}

\section{Introduction}
\label{subsection:sciencegoal}



The Fanaroff-Riley (FR) binary classification of radio galaxies has been widely adopted and utilised since its inception over four decades ago \cite{Fanaroff1974}. However, in spite of progress in relating the two classes to the dynamics and energetics of the sources \cite{Ineson2017,Saripalli2012}, our understanding of the causal relationship between the source physical properties/environment and FR classification is far from complete. To improve our inferences about the physics of these objects, we need to accurately identify and classify more radio galaxies while also detecting anomalous/rare examples \cite{Mingo2019}.

In particular for new sky surveys, such as those anticipated for the Square Kilometre Array (SKA) radio telescope, automated classification algorithms are increasingly being developed to replace the manual \textit{by eye} approaches used historically. CNNs have been used with success for image-based classification of radio galaxies (\cite{Wu2019, Aniyan2017}), including attempts to use more novel techniques such as capsule networks \cite{Lukic2018} and attention gating \cite{Bowles2021} to help improve performance and interpretability. \cite{Becker2020} provides a more comprehensive survey of current techniques. 

Currently, archival data-sets for training radio galaxy classifiers  are of comparable size to many of those used in computer vision (e.g. CIFAR \cite{Krizhevsky2009}), with around $10^5$ samples available. However, a fundamental difference is the sparsity of labels in radio galaxy data-sets: only a small fraction of data-points are labelled. This is largely due to the domain knowledge required for labelling, which incurs a high cost per label compared to typical machine learning data-sets. Currently, the largest labelled machine learning data-set of radio galaxies is  MiraBest \cite{Miraghaei2017, Porter2020MiraBestDataset} which has 1256 samples, orders of magnitude lower than the number of unlabelled images in its originating sky survey.

\begin{figure}
\centering
\begin{subfigure}[t]{.4\textwidth}
    \centering
    \includegraphics[width=0.9\linewidth]{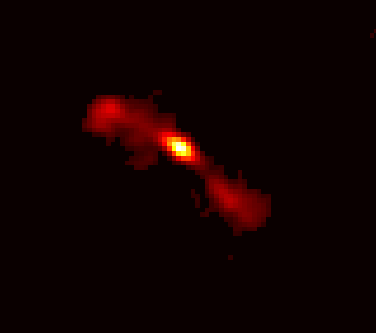}
    \caption{FR Class I}
\end{subfigure}
\begin{subfigure}[t]{.4\textwidth}
    \centering
    \includegraphics[width=0.9\linewidth]{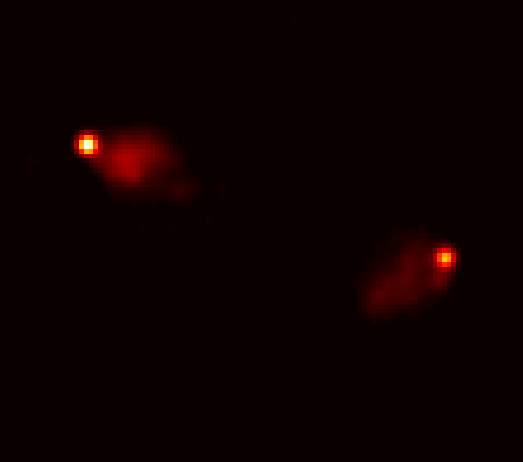}
    \caption{FR Class II}
\end{subfigure}
\caption{Example radio galaxies from the MiraBest dataset (cherry-picked for prototypicity) \cite{Miraghaei2017}.}
\label{figure:fr}
\end{figure}



\noindent
\textbf{This work:} In this work, we focus on morphological classification of radio-galaxies within the FR classification scheme, see Figure~\ref{figure:fr}. We aim to test whether we can achieve performance comparable to the current supervised state of the art on the MiraBest dataset with many fewer labels by using semi-supervised learning (SSL). This work will guide decision making on how to construct pipelines for upcoming radio surveys such as those for the SKA.

Many SSL algorithms are successful on benchmarking datasets such as CIFAR-10 \cite{Krizhevsky2009}, MNIST \cite{Lecun1998} and SVHN \cite{Netzer2011} in the regime of low data ($ \sim [10, 10^3]$ data points). However, less work has been done in assessing the robustness to various real world problems, such as unclean or covariate shifted data, new classes appearing in the unlabelled data or even simply varying sizes of labelled/unlabelled data: the shortcomings of the SSL literature have previously been documented in this context \cite{Oliver2018RealisticAlgorithms}.

Of particular interest in astronomy is the problem of biased sampling for our training data. We observe phenomena driven by complex natural processes which are selected for labelling in a biased manner due to instrumental, observational and intrinsic effects which favour, for example, particular flux density (brightness) and redshift (distance) ranges. This makes it difficult to know how representative our data catalogues are of all observed data, and the consequent effects on model performance are therefore not always clear a priori. Specific examples of how this is being addressed for machine learning applications in astronomy include the use of Gaussian process modelling to augment training data and make it more representative of test data in photometric classification \cite{Boone2019Avocado:Augmentation} and in galaxy merger classification where domain adaptation techniques have also been explored \cite{Ciprijanovic2020DomainMergers}. In both of these cases the solutions are tackling covariate shift between the labelled and test data.

\section{FixMatch}
\label{sec:fixmatch}
Semi-supervised learning leverages unlabelled data when only a small labelled data-set is available. The model is fed a set of image-label pairs $(\vx_l, y_l) \in \mX_l$ along with a (usually larger) set of unlabelled images $\vx_u \in \mX_u$. The goal is to predict labels for held out samples $\vx_{test} \in \mX_{test}$. An overview of semi-supervised learning can be found in \cite{ChapelleSSL}
\par
We choose the FixMatch algorithm \cite{Sohn2020FixMatch:Confidence} from the pool of SSL techniques as it achieves state of the art performance on benchmarking datasets, has few hyperparameters, and is computationally cheap \cite{Sohn2020FixMatch:Confidence}. FixMatch makes use of the unlabelled data through consistency regularisation and pseudo-labelling by adding a loss term computed on two different augmentations of the same image:
\begin{equation}
\label{equation:fmloss}
\mathcal{L} = \lambda \sum_{u=0}^{\mu B}\underbrace{\mathds{1}(\textup{max}(p_m(\alpha(\vx_u))) \geq \tau)}_\text{threshold mask} \times \underbrace{H(\hat{q}_u, p_m(y | \mathcal{A}(\vx_u)))}_\text{pseudo-label cross entropy} + 
\sum_{l=0}^B \underbrace{H(y_l , p_m(y | \alpha(\vx_l)))}_\text{supervised loss}.
\end{equation}
Here the "weak" augmentation, denoted $\alpha(\cdot)$, retains the semantic meaning of the image and simply uses the standard rotation/flipping augmentations. The "strong" augmentation, denoted by $\mathcal{A}( \cdot)$, uses RandAugment \cite{Cubuk2019} to apply a sequence of augmentations that can significantly alter the image.
\par
\begin{figure}[ht] 
    \centering
    \includegraphics[width=0.8\linewidth]{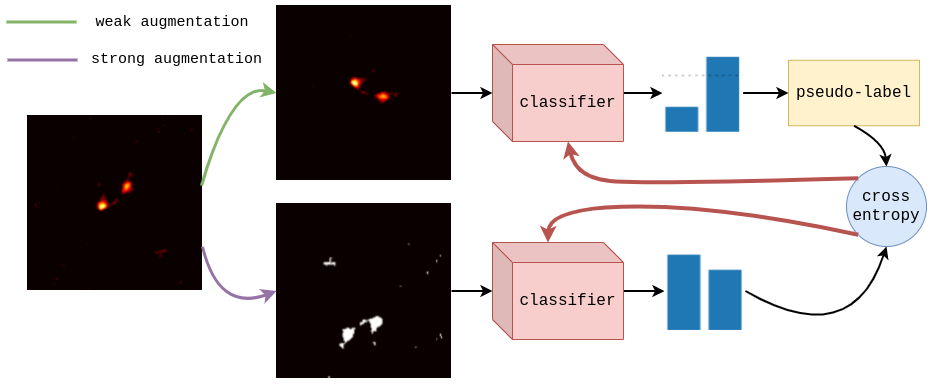}
    \caption{An unlabelled data point flowing through the FixMatch algorithm \cite{Sohn2020FixMatch:Confidence}}
    \label{figure:fixmatch}
\end{figure}

 
If $\mathrm{max}(p_m(y | \alpha(\vx_u))) > \tau$, the prediction $\hat{q}_u = \mathrm{argmax}(p_m(y | \alpha(\vx_u))$ is used as a pseudo-label for $\mathcal{A}(\vx_u)$. Figure~\ref{figure:fixmatch} shows how an unlabelled data point flows through the model. $B$ is the labelled batch size, $\mu$, $\tau$ are hyperparameters controlling unlabelled batch size and confidence threshold respectively. $\lambda$ controls the weighting of the unlabelled loss term and is set to 1, following \cite{Sohn2020FixMatch:Confidence}.

\section{Experiments}

\subsection{Ensuring fair evaluation}
\label{subsec:fairevaluation}
Fair evaluation is a problem in the SSL literature \cite{Oliver2018RealisticAlgorithms}, yet is crucial if we wish to apply it to real data. 
\textbf{Hyperparameters and validation set}: we keep the validation set a realistic size by scaling it to $20 \%$ of $\mX_l$, but set a hard lower limit to produce meaningful results when $\mX_l$ is small. We set the learning rate to 0.005 but scale the batch size, $B$, with $\mX_l$. Astronomical data-sets vary in size and content and large validation sets may not be available: our models need to perform well across a range of scenarios without too much hyperparameter sensitivity. For these reasons, we keep hyperparameter tuning to a minimum, opting instead to choose reasonable values close to the optimal values in the computer vision literature \cite{Sohn2020FixMatch:Confidence}. We choose $\mu = 7$, which allows the model to use a large proportion of the unlabelled data in a single epoch, and $\tau = 0.95$, which allows the model's confidence on unlabelled data to pass the threshold before it begins to overfit.
\textbf{Model architecture}: we use the convolutional architecture from \cite{Tang2019}, which has good performance for radio galaxy classification and relatively few parameters, giving a realistic baseline that isn't prone to overfitting. \textbf{Reproducibility and variance in results and data splits}: each experiment's results are averaged over 10 runs initialised using seeds 0-9. This ensures that the same splits are used during each experiment, while keeping weight initialisations consistent. We use randomly selected, stratified labelled/unlabelled data splits. The same test set is used after choosing the model weights with the best validation set accuracy.
\par
The experiments were performed on a single Nvidia A100 GPU with a total of 2.34 days of runtime. We used \href{https://wandb.ai/home}{Weights \& Biases} \cite{wandb} to track experiment results. Code can be found at \url{https://github.com/inigoval/fixmatch}.
    
\subsection{Does FixMatch outperform the baseline?}
\label{subsec:performance}

\textbf{Fully supervised baseline.} The baseline model is trained in a fully supervised fashion with a cross-entropy loss, shown as the second term in Equation~\ref{equation:fmloss}. We use the same subset of (weakly augmented) labelled samples as in the FixMatch case, but ignore the unlabelled samples.

\begin{figure}
\centering
\begin{subfigure}[t]{.45\textwidth}
    \centering
    \includegraphics[width=0.9\linewidth]{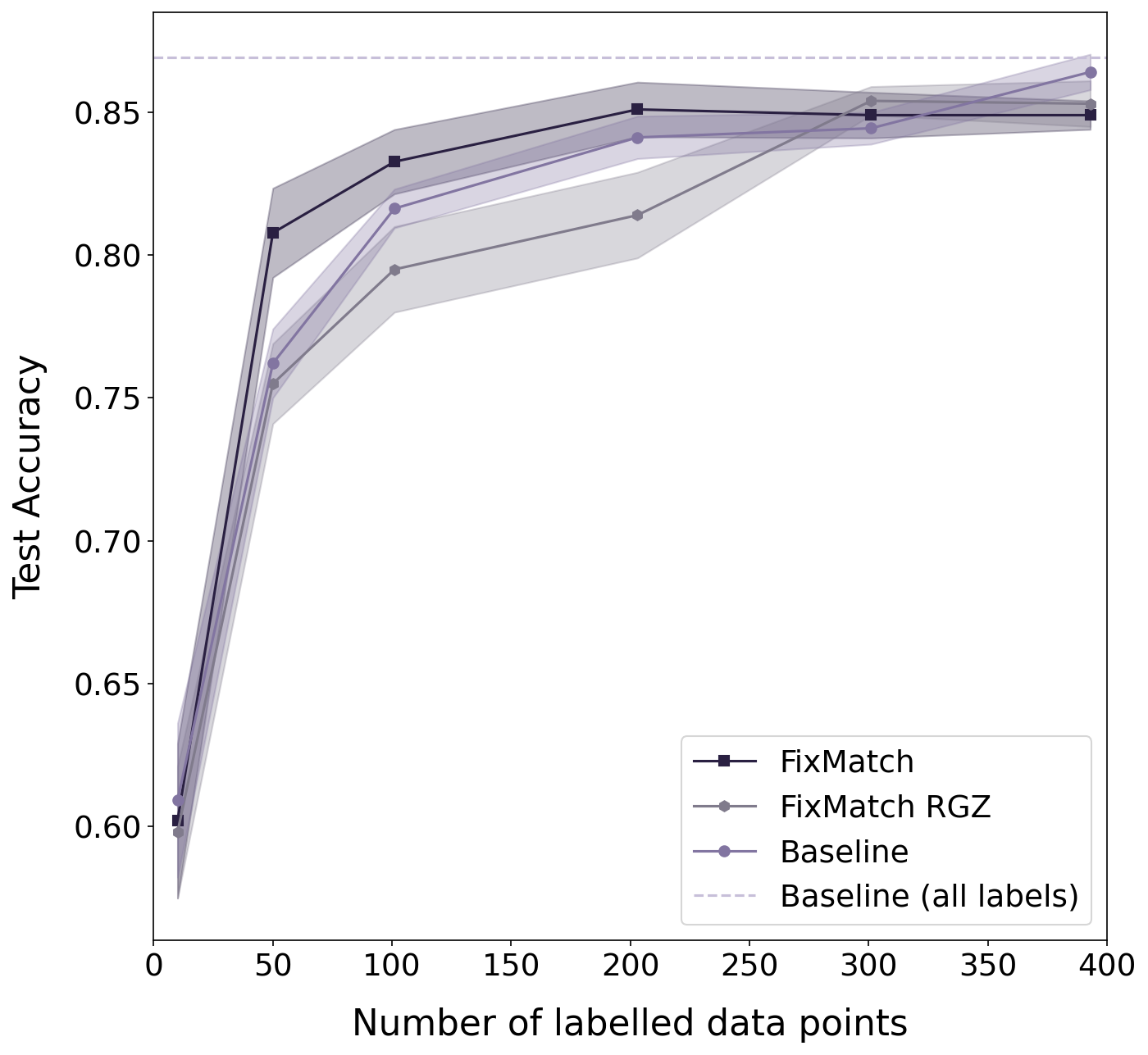}
    \caption{Accuracy on the test set.} 
    \label{fig:datafracacc}
\end{subfigure}
\begin{subfigure}[t]{.45\textwidth}
    \centering
    \includegraphics[width=0.9\linewidth]{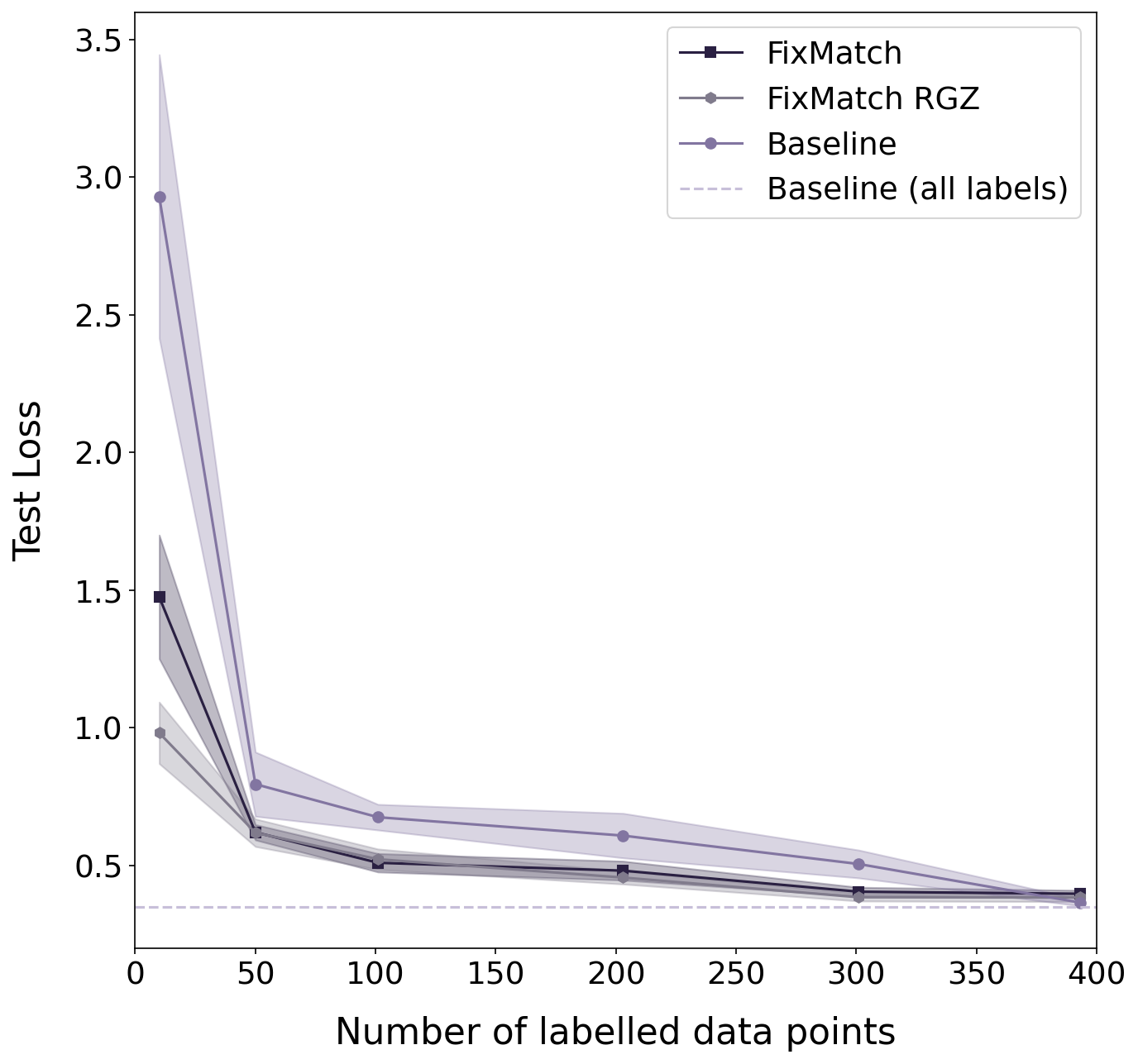}
    \caption{Loss on the test set} 
    \label{fig:datafracloss}
\end{subfigure}
\caption{Performance as a function of $\mX_l$ size.}
\label{figure:datafrac}
\end{figure}

\textbf{Throwing away labels to create an artificial SSL scenario (Case A)}. We perform multiple experiments by using a small labelled subset of the MiraBest dataset \cite{Miraghaei2017} and using the remainder as unlabelled data. In all cases we keep the labelled data stratified (class balance is preserved). In Figure~\ref{figure:datafrac} we see that FixMatch achieves a consistently lower loss on the test data and outperforms the baseline in test set accuracy when there is little labelled data available. We are able to recover comparable accuracy ($85.1 \% \pm 1.1\%$) to the supervised baseline with all labels ($86.93 \% \pm 0.54 \% $) using just $20\%$ (203) of the labels. However, FixMatch's performance degrades quickly with very few labels, which is in stark contrast to similar experiments in \cite{Sohn2020FixMatch:Confidence}, where good performance is achieved even with only one sample per class. Furthermore, the "sweet spot" where FixMatch has a significant advantage does not cover the full range of labelled data volumes.

\begin{figure}
\centering
\begin{subfigure}[t]{.45\textwidth}
\centering
\includegraphics[width=\linewidth]{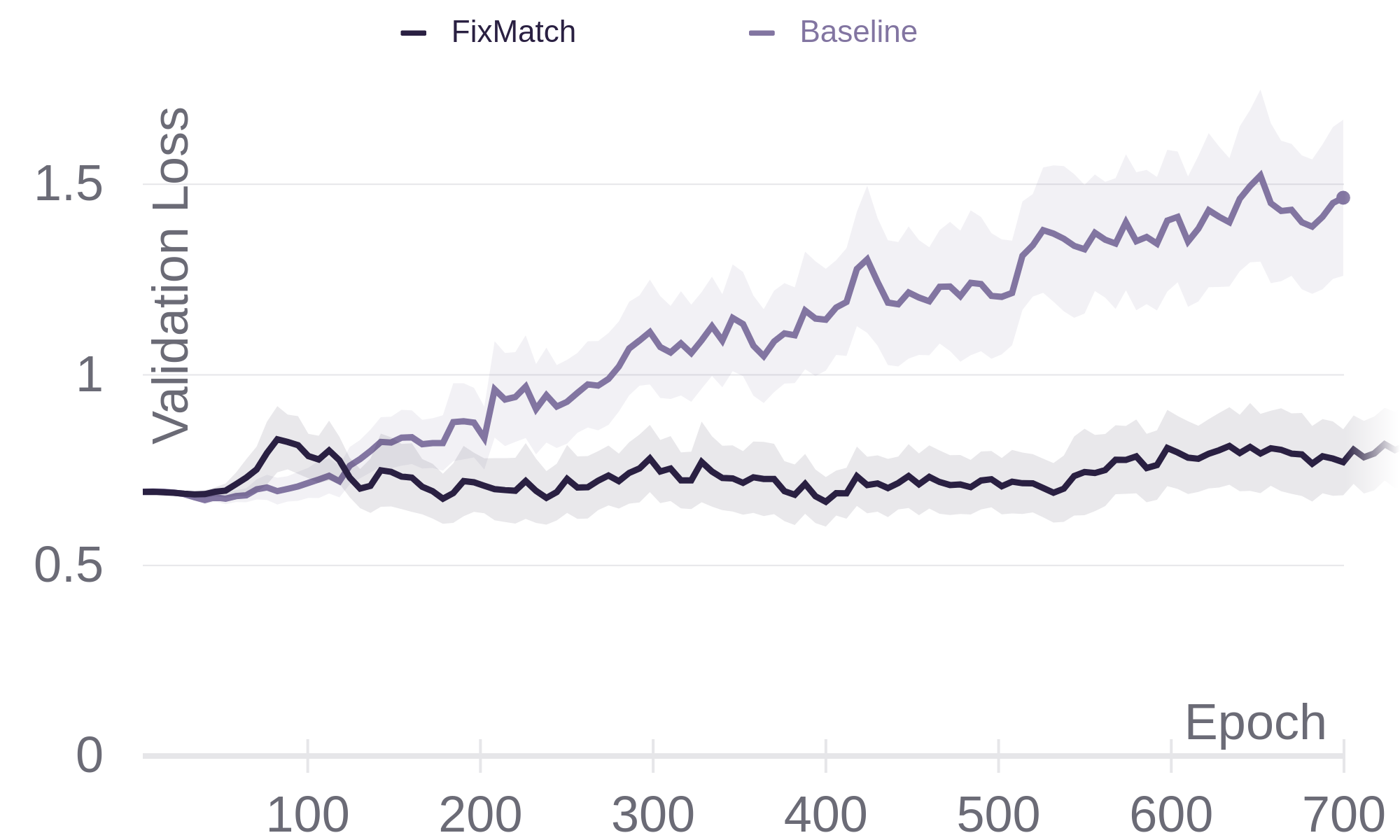}
\caption{Validation loss with 50 labelled samples.}
    \label{figure:val_loss0.05}
\end{subfigure}
\begin{subfigure}[t]{.45\textwidth}
    \centering
    \includegraphics[width=\linewidth]{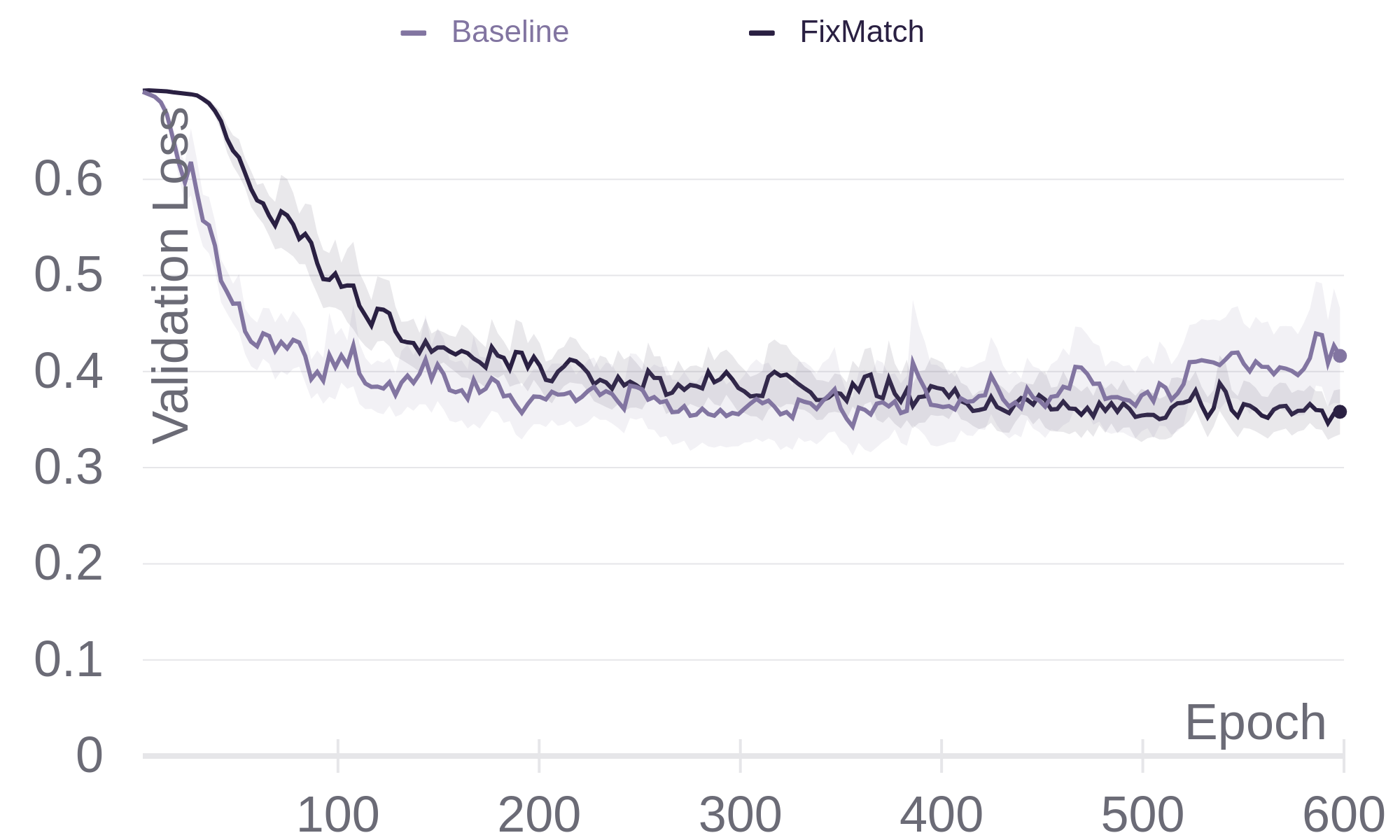}
    \caption{Validation loss with 393 labelled samples.} 
    \label{figure:val_loss0.4}
\end{subfigure}
\caption{FixMatch (Case A) regularisation effect on validation loss. The shaded area shows the standard error over 10 runs. We use exponential moving average smoothing with a weight of 0.3.}
\label{figure:val_loss}
\end{figure}

We hypothesise that there is a strong regularisation effect from the strongly augmented samples, as well as new information being learned from the unlabelled data. This is illustrated in Figure~\ref{figure:val_loss0.05}, where FixMatch avoids overfitting at low data volumes, as demonstrated by the well behaved validation loss. We also observe that this effect is not present with 393 labelled samples, which is reflected in the equal test loss at high label volumes in Figure~\ref{figure:val_loss0.4}.

\label{section:datatest}

\textbf{Testing FixMatch on real unlabelled data (Case B)}. We test FixMatch by using a pool of 20,000 unlabelled data from the Radio Galaxy Zoo Data Release 1 (RGZ DR1) catalogue (Wong et al. in prep) with labelled samples from MiraBest. While these data originate from the same radio survey and are pre-processed in the same manner as MiraBest, the class balance of the RGZ DR1 catalogue is unknown and the choice of filters for choosing data-points is wider, resulting in a dataset of $\sim 10^5$ datapoints. We find that the RGZ DR1 data-set contains many unresolved sources, which are uninformative to our model. To remove these sources, we enforce a lower limit on source extension by calculating the Frechet Distance (FD; \cite{Heusel2017}) between the labelled data-set and the RGZ DR1 dataset (in feature space) for a range of lower limits. We take the limit with the lowest score (28 arcsec), after which a total of 8848 unlabelled data samples remain.

We find that in Case B FixMatch still provides regularisation as shown by the comparable test set loss to Case A, see Figure~\ref{fig:datafracloss}. However, this does not result in an improvement in accuracy. Indeed we  see a decrease in test set accuracy compared to the baseline in Case B, for all labelled data volumes. We hypothesise that this may be due to the unknown class imbalance in the RGZ DR1 data-set.

\begin{figure}
\centering
\begin{subfigure}[t]{.45\textwidth}
\centering
\includegraphics[width=\linewidth]{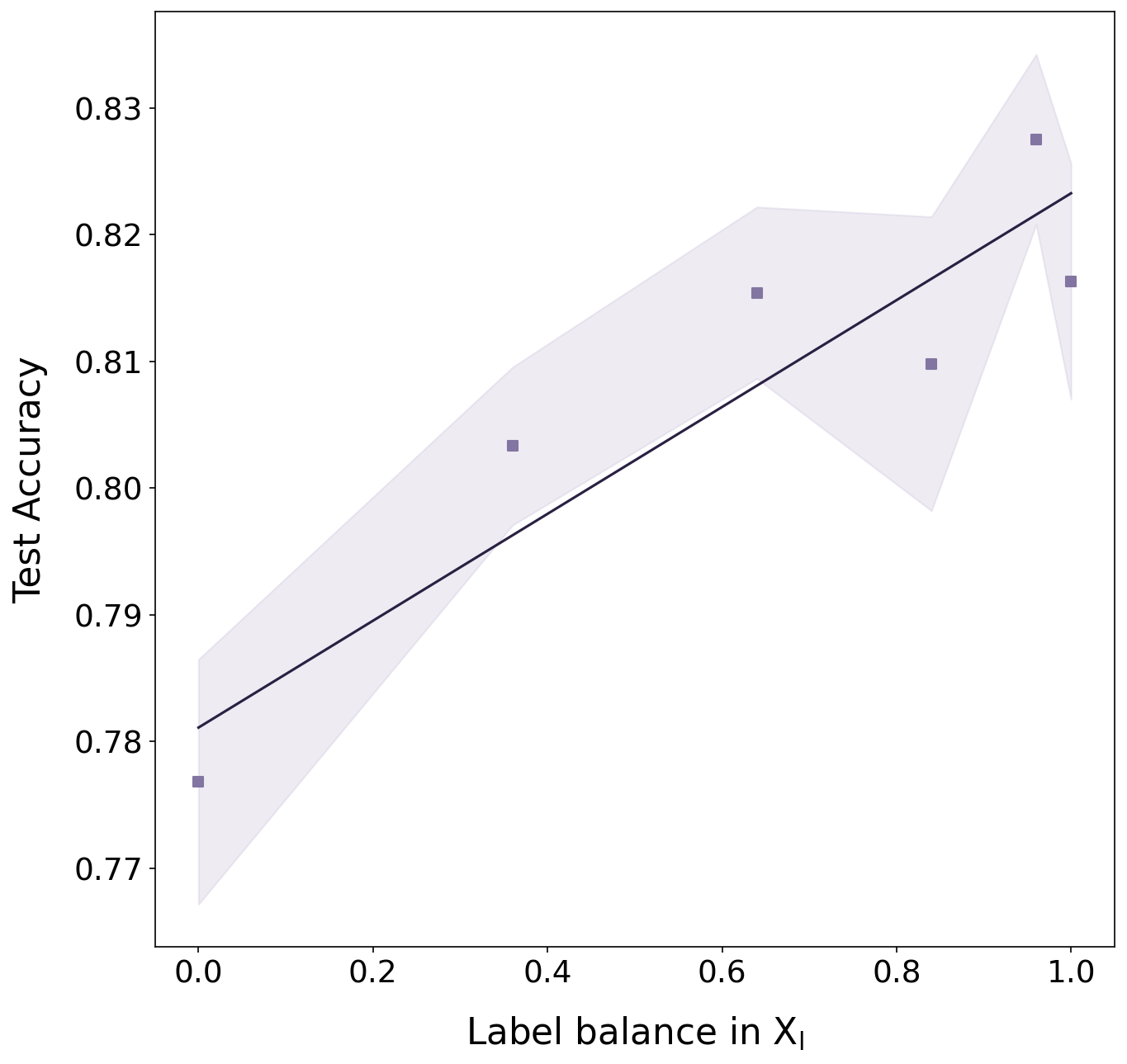}
\caption{Removing FRI samples}
    \label{fig:balaccfri}
\end{subfigure}
\begin{subfigure}[t]{.45\textwidth}
    \centering
    \includegraphics[width=\linewidth]{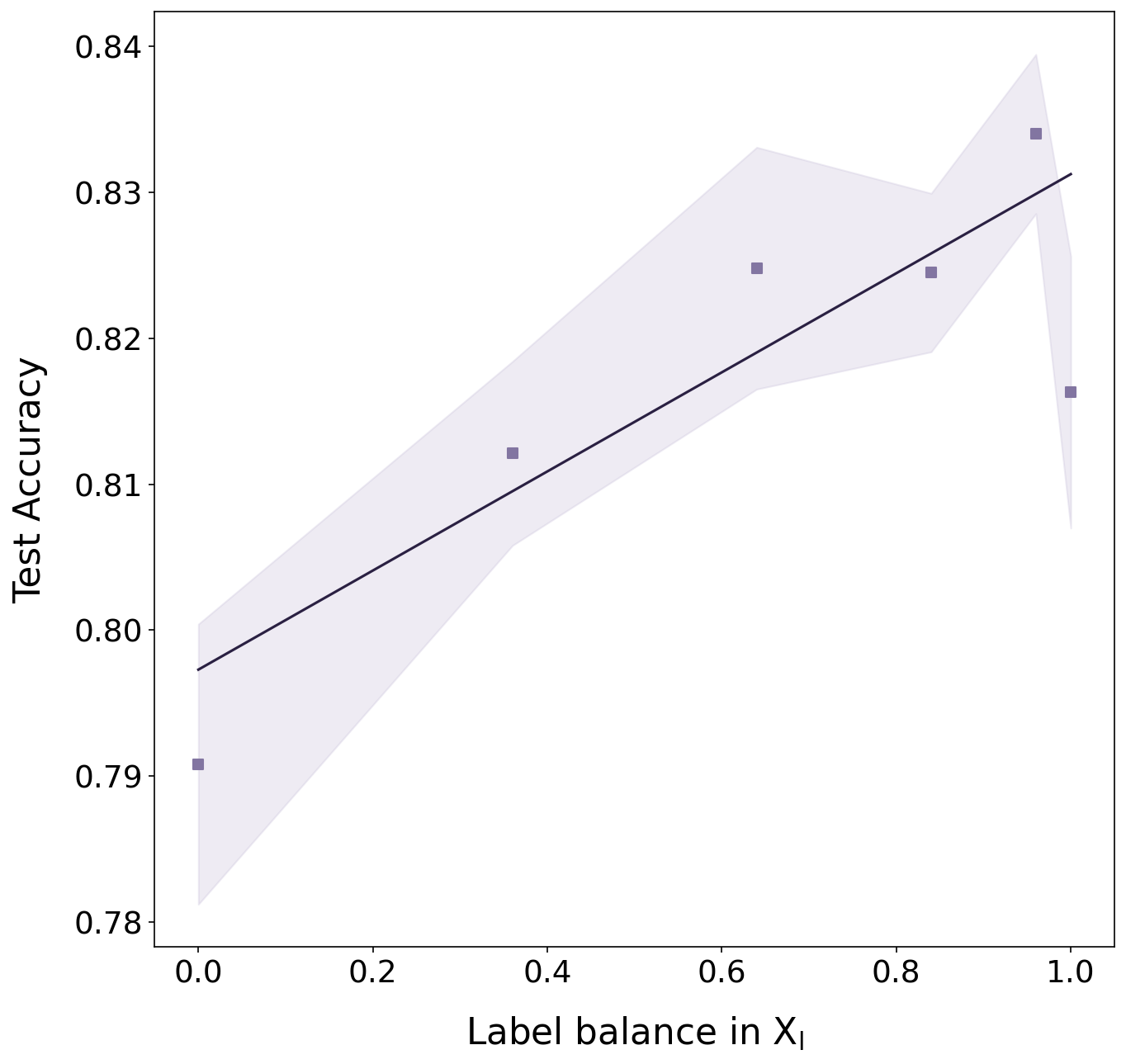}
    \caption{Removing FRII samples.} 
    \label{fig:balaccfrii}
\end{subfigure}
\caption{Test set accuracy as a function of label balance in the unlabelled (MiraBest) data-set. We use only samples qualified as confidently classified in the MiraBest dataset for the labelled subset (69 labels) to reduce the noise in our results, see \cite{Scaife2021} for more details on this qualification. Label balance is quantified by computing $4(1-\beta) \beta $. Error bars show the standard error after aggregating 20 runs (seeded 0-19).}
\label{fig:balacc}
\end{figure}

To test this hypothesis we rerun Case A with an artificially class imbalanced unlabelled data-set, produced by removing either FRI or FRII samples from the unlabelled data.
We measure test set accuracy for unlabelled data with varying proportion $\beta$ of FRIs. Figure~\ref{fig:balacc} shows that class imbalance has a significant negative effect on test set accuracy. This suggests that the performance gap between Cases A and B could be accounted for by class imbalance in RGZ DR1, although it is not conclusive that this is the only factor.

    
\section{Conclusion}
We find that FixMatch provides some regularisation benefits when learning with few labelled samples, mitigating the effect of overfitting, as well as learning from unlabelled data. Furthermore, we are able to achieve better accuracy on the test set than the baseline with fewer labels. While this is relatively promising, the improvement in accuracy is much smaller than for the computer vision data-sets the algorithm was initially tested on. 

Poor results using the ``real'' RGZ DR1 data highlight an important obstacle to applying SSL ``in the wild'' on scientific observational data: $\mX_l$ and $\mX_u$ are unlikely to be drawn from identical distributions. We see that even class imbalance in $\mX_{\rm u}$ can have a major effect on SSL performance, noting that in a real scenario we have no way of knowing the class balance of $\mX_{\rm u}$.

We believe that a naive application of SSL, although it may outperform the baseline in some cases and provide useful regularisation, requires further domain specific development to give a worthwhile advantage in the case of radio galaxy morphology classification. Future work will consider handcrafting radio galaxy specific augmentations and implementing more sophisticated SSL approaches to help tackle the problem of covariate shift between $\mX_{\rm u}$ and $\mX_{\rm l}$, such as pretraining and/or incorporating domain adaptation (e.g. \cite{Cai2021}).

\begin{ack}
The authors gratefully acknowledge support from the UK Alan Turing Institute under grant reference EP/V030302/1. IVS gratefully acknowledges support from the Frankopan Foundation. 

This work has been made possible by the participation of more than 12,000 volunteers in the Radio Galaxy Zoo Project. The data in this paper are the result of the efforts of the Radio Galaxy Zoo volunteers, without whom none of this work would be possible. Their efforts are individually acknowledged at \url{ http://rgzauthors.galaxyzoo.org}.
\end{ack}

\clearpage

\bibliography{references}
\bibliographystyle{abbrv}

\clearpage

\end{document}

%% file: math_commands.tex
\usepackage{amsmath,amsfonts,bm}

\def\eqref#1{equation~\ref{#1}}

\def\1{\bm{1}}

\def\vx{{\bm{x}}}

\def\mX{{\bm{X}}}

\DeclareMathAlphabet{\mathsfit}{\encodingdefault}{\sfdefault}{m}{sl}
\SetMathAlphabet{\mathsfit}{bold}{\encodingdefault}{\sfdefault}{bx}{n}

